\documentclass[aps,rpl,preprint,groupedaddress]{revtex4}
\begin{document}

\title{Optimal approximate reversal of quantum operations on a single qubit}
\author{Ye Yeo}

\affiliation{Centre for Mathematical Sciences, Wilberforce Road, Cambridge CB3 0WB, United Kingdom}

\begin{abstract}
We demonstrate how insights gained from reformulating the problem of quantum teleportation into one of reversing quantum operations, and designing optimum completely positive maps for teleportation, can enable one to explore optimal approximate reversal of quantum operations on a single qubit.  In particular, we show that the optimal approximate reversal of a generalized depolarizing channel can be achieved using only unitary transformations.  We also show that for a quantum channel, which reveals some information about the input state, extremal completely positive maps and not unitary transformations yield optimal approximate reversal.
\end{abstract}

\maketitle

The formalism of quantum operations, described in detail by Kraus \cite{Kraus}, describes the most general possible state change in quantum mechanics.  In this formalism there is an input state $\rho_{in}$ and an output state $\rho_{out}$, which are connected by a map
$$
\rho_{in} \longrightarrow \rho_{out} 
= \frac{{\cal E}(\rho_{in})}{{\rm tr}[{\cal E}(\rho_{in})]}.
$$
The map is determined by a quantum operation $\cal E$, a linear, trace-decreasing map that preserves complete positivity.  The most general form for $\cal E$ can be shown to be \cite{Kraus}
$$
{\cal E}(\rho_{in}) = \sum_k A_k\rho_{in}A^{\dagger}_k,
$$
\begin{equation}
\sum_k A^{\dagger}_kA_k \leq I.
\end{equation}
The Kraus operators $A_k$ completely specify the quantum operation $\cal E$.  In the special case of unitary evolution $U$ experienced by a closed quantum system, there is only one term in the operator-sum representation (1), $A_1 = U$, leaving us with
$$
\rho_{in} \longrightarrow \rho_{out} = U\rho_{in}U^{\dagger}.
$$
Unitary evolution is invertible, that is, the original input state $\rho_{in}$ can be recovered with certainty by subjecting the output state $\rho_{out}$ to the inverse unitary evolution $U^{-1}$.  This is not true for state change occuring in open quantum systems.  The reversal of general quantum operations therefore becomes an important issue in quantum information processing (see Ref. \cite{Caves} and references therein).  A quantum operation $\cal E$ is ``perfectly'' reversible on a subspace $\cal G$ of the total state space $\cal H$ if there exists a trace-preserving quantum operation $\cal R$, acting on the total state space $\cal H$, such that for all input states $\rho_{in}$ whose support lies in $\cal G$, \cite{Caves}
$$
\rho_{in} =
{\cal R}\left(\frac{{\cal E}(\rho_{in})}{{\rm tr}[{\cal E}(\rho_{in})]}\right) 
= \frac{{\cal R}\circ{\cal E}(\rho_{in})}{{\rm tr}[{\cal E}(\rho_{in})]}.
$$
The necessary and sufficient conditions for a general quantum operation to be perfectly reversible were derived in Ref. \cite{Caves}.

In this paper, we consider the problem of optimal approximate reversal of a quantum operation $\cal E$ on the whole state space $\cal H$ instead.  From hereon, reversal means approximate reversal.  We show how the techniques used in Ref. \cite{Rehacek} to identify and design optimum completely positive maps for quantum teleportation, can be employed to tackle our problem.  The optimality of a given $\cal R$ will be judged by the fidelity \cite{Jozsa} $F(\rho_{in}, {\cal R}\circ{\cal E}(\rho_{in}))$ between $\rho_{in}$ and ${\cal R}\circ{\cal E}(\rho_{in})$ averaged over an isotropic distribution of input states $\rho_{in}$.  We show explicitly that the action of a generalized depolarizing channel on a single qubit can be reversed optimally using only unitary transformations.  We also consider a quantum operation which reveals information about the input state.  In this case, optimal reversal requires more general quantum operations which are not unitary.

We begin with the observation \cite{Nielsen} that one can recast the problem of achieving optimal quantum teleportation \cite{Bennett} into one of optimal reversal of quantum operations.  A general teleportation scheme involves a sender, Alice, and a receiver, Bob.  Alice is in possession of two $n$-level quantum systems, the input system $1$, and another system $2$ arbitrarily entangled with a third $n$-level target system $3$ in Bob's possession.

Initially the composite system $123$ is prepared in a state with density operator $\tilde{\rho}_1 \otimes \chi_{23}$, where $\tilde{\rho}_1$ is an unknown state of the input system $1$, and $\chi_{23}$ is an arbitrary entangled state of systems $2$ and $3$.  Since the systems $1$ and $3$ are identical and thus have the same state space, a one-to-one correspondence from the state space of the composite system onto itself can be established by a unitary swap operator $U_{1(2)3}$, which acts on product states according to
$$
U_{1(2)3}(|\tilde{a}\rangle_1 \otimes |b\rangle_2 \otimes |c\rangle_3) =
|\tilde{c}\rangle_1 \otimes |b\rangle_2 \otimes |a\rangle_3,
$$
swapping the states of systems $1$ and $3$, while leaving system $2$ alone.  $U_{1(2)3}$ obviously satisfies $(U_{1(2)3})^2 = I_{123}$, the identity operator on the composite system, and $U^{\dagger}_{1(2)3} = U_{1(2)3}$.  When extended to operators $Q_{123}$ on the composite system, the correspondence becomes
$$
\tilde{Q}_{123} \leftrightarrow Q_{123} 
= U_{1(2)3}\tilde{Q}_{123}U^{\dagger}_{1(2)3}.
$$
It follows that
\begin{equation}
\tilde{\rho}_1 \otimes \chi_{23} = U_{1(2)3}(\tilde{\chi}_{12} \otimes \rho_3)
U^{\dagger}_{1(2)3},
\end{equation}
where $\tilde{\chi}_{12}$ is the counterpart of $\chi_{23}$.

To teleport the input state $\tilde{\rho}_1$ to Bob's target system $3$, Alice performs a generalized measurement on systems $1$ and $2$.  This generalized measurement is described by operators $\tilde{\Pi}^{ij}_{12} \otimes I_3$, where $\tilde{\Pi}^{ij}_{12}$ are Kraus operators on the joint system $12$, $i$ labels the outcome of the measurement, and
$$
\sum_i\sum_j \tilde{\Pi}^{ij\dagger}_{12}\tilde{\Pi}^{ij}_{12} = I_{12}.
$$
If Alice's measurement has outcome $i$, she communicates her measurement result to Bob via a classical channel.

The state of Bob's target system $3$ conditioned on Alice's measurement result $i$ is given by
\begin{equation}
\rho^i_3 = \frac{1}{p_i}{\rm tr}_{12}\left[
\sum_j (\tilde{\Pi}^{ij}_{12} \otimes I_3)(\tilde{\rho}_1 \otimes \chi_{23})
(\tilde{\Pi}^{ij\dagger}_{12} \otimes I_3)
\right],
\end{equation}
where
$$
p_i = {\rm tr}_{123}\left[
\sum_j (\tilde{\Pi}^{ij}_{12} \otimes I_3)(\tilde{\rho}_1 \otimes \chi_{23})
(\tilde{\Pi}^{ij\dagger}_{12} \otimes I_3)
\right].
$$
Substituting Eq.(2) into Eq.(3) gives
\begin{equation}
\rho^i_3 = \frac{1}{p_i}{\rm tr}_{12}\left[
\sum_j (\tilde{\Pi}^{ij}_{12} \otimes I_3)U_{1(2)3}
(\tilde{\chi}_{12} \otimes \rho_3)
U^{\dagger}_{1(2)3}(\tilde{\Pi}^{ij\dagger}_{12} \otimes I_3)
\right].
\end{equation}
Writing
$$
\tilde{\chi}_{12} = \sum_k q_k|\tilde{s}_k\rangle_{12}\langle\tilde{s}_k|,
$$
where the vectors $|\tilde{s}_k\rangle_{12}$ make up the complete orthonormal set of eigenvectors of $\tilde{\chi}_{12}$ in the joint space of $1$ and $2$, and performing the partial trace of Eq.(4) in the complete orthonormal basis $|\tilde{P}_l\rangle_{12}$ for the joint system $12$ gives
$$
\rho^i_3 = \frac{1}{p_i}\sum_{j, k, l}q_k\ 
{_{12}}\langle\tilde{P}_l|
(\tilde{\Pi}^{ij}_{12} \otimes I_3)U_{1(2)3}
(|\tilde{s}_k\rangle_{12}\langle\tilde{s}_k| \otimes \rho_3)
U^{\dagger}_{1(2)3}(\tilde{\Pi}^{ij\dagger}_{12} \otimes I_3)
|\tilde{P}_l\rangle_{12}
$$
$$
= \sum_{j, k, l}\left[\sqrt{\frac{q_k}{p_i}}
{_{12}}\langle\tilde{P}_l|(\tilde{\Pi}^{ij}_{12} \otimes I_3)U_{1(2)3}
|\tilde{s}_k\rangle_{12}\right]\rho_3\left[
\sqrt{\frac{q_k}{p_i}}
{_{12}}\langle\tilde{s}_k|U^{\dagger}_{1(2)3}
(\tilde{\Pi}^{ij\dagger}_{12} \otimes I_3)
|\tilde{P}_l\rangle_{12}
\right].
$$
Therefore, $\rho^i_3$ is related to $\rho_3$ by a quantum operation ${\cal E}^i$:
\begin{equation}
\rho^i_3 = {\cal E}^i(\rho_3) = \sum_m A^{im}_3\rho_3A^{im\dagger}_3,
\end{equation}
where
$$
A^{im}_3 \equiv \sqrt{\frac{q_k}{p_i}}
{_{12}}\langle\tilde{P}_l|(\tilde{\Pi}^{ij}_{12} \otimes I_3)U_{1(2)3}
|\tilde{s}_k\rangle_{12},
$$
\begin{equation}
\sum_m A^{im\dagger}_3A^{im}_3 \leq I_3,
\end{equation}
and the single index $m$ denotes the triple $(j, k, l)$.

For Bob to successfully complete the teleportation protocol, he must perform a $i$-dependent trace-preserving quantum operation ${\cal R}^i$:
$$
{\cal R}^i(\rho^i_3) = \sum_n B^{in}_3\rho^i_3B^{in\dagger}_3,
$$
\begin{equation}
\sum_n B^{in\dagger}_3B^{in}_3 = I_3,
\end{equation}
such that the fidelity \cite{Jozsa} $F^i(\rho_3, {\cal R}^i\circ{\cal E}^i(\rho_3))$ between the input state $\rho_3$ and the teleported state ${\cal R}^i\circ{\cal E}^i(\rho_3)$ is optimal, that is, as close to one as possible.  In other words, Bob has to determine ${\cal R}^i$ which optimally reverses ${\cal E}^i$.

Before we discuss how optimal ${\cal R}^i$ can be determined, we demonstrate that for two-level systems $1$, $2$ and $3$ (from hereon we consider only two-level systems), with
\begin{equation}
\chi_{23} = q_1|\Phi^+\rangle_{23}\langle\Phi^+| +
q_2|\Phi^-\rangle_{23}\langle\Phi^-| +
q_3|\Psi^+\rangle_{23}\langle\Psi^+| +
q_4|\Psi^-\rangle_{23}\langle\Psi^-|
\end{equation}
where $0 \leq q_k \leq 1$, $\sum^4_{k = 1}q_k = 1$,
$$
|\Phi^{\pm}\rangle_{23} = \frac{1}{\sqrt{2}}(|00\rangle_{23} \pm |11\rangle_{23}),
$$
$$
|\Psi^{\pm}\rangle_{23} = \frac{1}{\sqrt{2}}(|01\rangle_{23} \pm |10\rangle_{23}),
$$
are the Bell states, and
$$
\tilde{\Pi}^{1j}_{12} = \tilde{\Pi}^1_{12} = 
|\tilde{\Phi}^+\rangle_{12}\langle\tilde{\Phi}^+|,\
\tilde{\Pi}^{2j}_{12} = \tilde{\Pi}^2_{12} =
|\tilde{\Phi}^-\rangle_{12}\langle\tilde{\Phi}^-|,
$$
\begin{equation}
\tilde{\Pi}^{3j}_{12} = \tilde{\Pi}^3_{12} = 
|\tilde{\Psi}^+\rangle_{12}\langle\tilde{\Psi}^+|,\
\tilde{\Pi}^{4j}_{12} = \tilde{\Pi}^4_{12} =
|\tilde{\Psi}^-\rangle_{12}\langle\tilde{\Psi}^-|,
\end{equation}
then ${\cal E}^1$ is a generalized depolarizing channel.  Here, we use $|0\rangle$ and $1\rangle$ to denote an orthonormal set of basis states for each two-level system.  Eq.(8) and Eq.(9) allow us to calculate $p_1 = p_2 = p_3 = p_4 = \frac{1}{4}$.  Substituting Eq.(8) into Eq.(2), we obtain
$$
\tilde{\chi}_{12} = q_1|\tilde{\Phi}^+\rangle_{12}\langle\tilde{\Phi}^+| +
q_2|\tilde{\Phi}^-\rangle_{12}\langle\tilde{\Phi}^-| +
q_3|\tilde{\Psi}^+\rangle_{12}\langle\tilde{\Psi}^+| +
q_4|\tilde{\Psi}^-\rangle_{12}\langle\tilde{\Psi}^-|.
$$
That is, the complete orthonormal set of eigenvectors of $\tilde{\chi}_{12}$ in the joint space of $1$ and $2$ is given by
\begin{equation}
|\tilde{s}_1\rangle_{12} = |\tilde{\Phi}^+\rangle_{12},\
|\tilde{s}_2\rangle_{12} = |\tilde{\Phi}^-\rangle_{12},\
|\tilde{s}_3\rangle_{12} = |\tilde{\Psi}^+\rangle_{12},\
|\tilde{s}_4\rangle_{12} = |\tilde{\Psi}^-\rangle_{12}.
\end{equation}
Eq.(6), when $i = 1$, is in this case reduced to
\begin{equation}
A^{1kl}_3 = \sqrt{\frac{q_k}{p_1}}
{_{12}}\langle\tilde{P}_l|(\tilde{\Pi}^1_{12} \otimes I_3)U_{1(2)3}
|\tilde{s}_k\rangle_{12}.
\end{equation}
Making the choice
\begin{equation}
|\tilde{P}_1\rangle_{12} = |\tilde{\Phi}^+\rangle_{12},\
|\tilde{P}_2\rangle_{12} = |\tilde{\Phi}^-\rangle_{12},\
|\tilde{P}_3\rangle_{12} = |\tilde{\Psi}^+\rangle_{12},\
|\tilde{P}_4\rangle_{12} = |\tilde{\Psi}^-\rangle_{12},
\end{equation}
and substituting Eq.(10) and Eq.(12) into Eq.(11) yields a generalized depolarizing channel ${\cal E}^1$ specified by Kraus operators
\begin{equation}
A^{111}_3 = \sqrt{q_1}I_3,\
A^{121}_3 = \sqrt{q_2}\sigma^z_3,\
A^{131}_3 = \sqrt{q_3}\sigma^x_3,\
A^{141}_3 = \sqrt{q_4}\sigma^y_3,
\end{equation}
where
$$
\sigma^x = \left(\begin{array}{cc} 0 & 1 \\ 1 & 0 \end{array}\right),\
\sigma^y = \left(\begin{array}{cc} 0 & -i\\ i & 0 \end{array}\right),\
\sigma^z = \left(\begin{array}{cc} 1 & 0 \\ 0 & -1\end{array}\right)
$$
are the Pauli matrices.

Recently, Rehacek {\it et al.} \cite{Rehacek} found an efficient iterative algorithm for identifying quantum operations ${\cal R}^i$ on Bob's side which optimizes the average teleportation fidelity $\langle F\rangle$.  In particular, for two-level systems $1$, $2$ and $3$, the connection between optimum ${\cal R}^i$ and extremal completely positive maps was discussed.  Assume that the input states are pure, that is,
$$
\tilde{\rho}_1 = |\tilde{\psi}\rangle_1\langle\tilde{\psi}|,\
|\tilde{\psi}\rangle_1 = \cos\frac{\theta}{2}|0\rangle_1 + e^{i\phi}\sin\frac{\theta}{2}|1\rangle_1,
$$
where $\theta$ and $\phi$ are the polar and azimuthal angles respectively.  No generality is lost by this assumption, since mixed states are just convex combinations of pure states.  $\langle F\rangle$ is obtained by averaging over all possible Alice's measurement outcomes $i$ and over an isotropic distribution of input states $\tilde{\rho}_1$:
$$
\langle F\rangle = \sum_i p_i\int d\psi 
F^i(\rho_3, {\cal R}^i\circ{\cal E}^i(\rho_3))
$$
$$
= \frac{1}{4\pi}\sum_i\sum_n{\rm tr}_{123}\left[
\int^{\pi}_0\int^{2\pi}_0 \sin\theta d\theta d\phi\
\rho_3B^{in}_3(
\sum_j\tilde{\Pi}^{ij\dagger}_{12}\tilde{\Pi}^{ij}_{12} \otimes I_3)
(\tilde{\rho}_1 \otimes \chi_{23})B^{in\dagger}_3
\right]
$$
\begin{equation}
= \frac{1}{2} + \frac{1}{12}\sum_i{\rm tr}_3(\vec{X}^i\cdot\vec{T}^i).
\end{equation}
Here \cite{Rehacek},
\begin{equation}
\vec{X}^i \equiv \sum_n B^{in\dagger}_3\vec{\sigma}_3B^{in}_3 = M\vec{\sigma}_3 + \vec{c},
\end{equation}
where $M$ is a $3 \times 3$ matrix, $\vec{c}$ is a column $3$-vector, and
\begin{equation}
\vec{T}^i \equiv {\rm tr}_{12}
[(\sum_j\tilde{\Pi}^{ij\dagger}_{12}\tilde{\Pi}^{ij}_{12} \otimes I_3)(\vec{\sigma}_1 \otimes \chi_{23})].
\end{equation}
Two observations about Eq.(14) are important.  First, quantum operations ${\cal R}^i$ corresponding to different Alice's measurement outcome $i$ are independent, therefore each term on the right-hand side of Eq.(14) can be maximized independently.  In particular, when we consider the case above, for $i = 1$, we will be looking for the quantum operation ${\cal R}^1$ which optimally reverses the generalized depolarizing channel (13).  Second, $\langle F\rangle$ is a linear functional of $\vec{X}^i$.  This means that all its maxima lie on the boundary of the set of physically allowed operators $\vec{X}^i$ that is determined by the constraint of complete positivity of the corresponding transformations.  The set of extremal completely positive maps comprising the boundary of the convex set of all completely positive maps thus contains all Bob's optimum transformations.  The matrix $M$ can always be brought to a diagonal form via a unitary transformation.  When in diagonal form, extremal completely positive maps can be parametrized by two angles $u \in [0,\ 2\pi)$ and $v \in [0,\ \pi)$, \cite{Rehacek}
$$
M = \left(\begin{array}{ccc}
\cos u & 0 & 0 \\
0 & \cos v & 0 \\
0 & 0 & \cos u\cos v
\end{array}\right),\
\vec{c} = \left(\begin{array}{c}
0 \\ 0 \\ \sin u\sin v
\end{array}\right).
$$

Substituting $\tilde{\Pi}^1_{12} = |\tilde{\Phi}^+\rangle_{12}\langle\tilde{\Phi}^+|$ and Eq.(8) into Eq.(16), we obtain
\begin{equation}
T^1_x = \frac{1}{4}t_x\sigma^x_3,\
T^1_y = \frac{1}{4}t_y\sigma^y_3,\
T^1_z = \frac{1}{4}t_z\sigma^z_3,
\end{equation}
where
$$
t_x = q_1 - q_2 + q_3 - q_4,
$$
$$
t_y = q_1 - q_2 - q_3 + q_4,
$$
$$
t_z = q_1 + q_2 - q_3 - q_4.
$$
Note that $-1 \leq t_x, t_y, t_z \leq 1$.  The fidelity between a pure input state $\rho_{in}$ and the output state ${\cal R}^1\circ{\cal E}^1(\rho_{in})$, which has gone through the generalized depolarizing channel ${\cal E}^1$ followed by a ``reversal channel'' ${\cal R}^1$, averaged over an isotropic distribution of input states is thus given by
$$
\langle F^1\rangle = \int d\psi F^1(\rho_{in}, {\cal R}^1\circ{\cal E}^1(\rho_{in}))
$$
\begin{equation}
= \frac{1}{2} + \frac{1}{6}(t_x\cos u + t_y \cos v + t_z\cos u\cos v).
\end{equation}

For $t_x, t_y, t_z = 0$, we have a totally random channel and the optimum $\langle F^1\rangle$ is expectedly one-half.  If $0 < t_x, t_y, t_z \leq 1$, then $\cos u = \cos v = 1$ optimizes $\langle F^1\rangle$.  That is, the optimum reversal channel is the ``unitary'' noiseless channel specified by $B^1 = I$.  The noiseless, bit flip, phase flip, bit-phase flip, and depolarizing channels, with $q_1 > q_2, q_3, q_4$, belong to this category.  Another category is when any two elements from the set $\{t_x, t_y, t_z\}$ are negative but the third element is positive, then the optimum reversal channel is the ``unitary'' channel specified by either $B^1 = \sigma^x$, or $B^1 = \sigma^y$, or $B^1 = \sigma^z$.  For instance, $q_2 = 1$ or $t_x = t_y = -1, t_z = 1$, gives the unitary phase flip channel, which can be optimally reversed by $B^1 = \sigma^z$.  For these two categories, we have
\begin{equation}
\langle F^1_{\max}\rangle = \frac{1}{2} + \frac{1}{6}(\max\{
4q_1 - 1, 4q_2 - 1, 4q_3 - 1, 4q_4 - 1\})
\end{equation}

When only one element from the set $\{t_x, t_y, t_z\}$ is negative or when $-1 \leq t_x, t_y, t_z < 0$, the unitary reversal channels yield Eq.(19).  In these cases, for $|t_x|, |t_y| < |t_z|$, we could use more general ``nonunitary'' trace-preserving reversal channels which optimize $\langle F^1\rangle$.  These yield
\begin{equation}
\langle F^1_{optimal}\rangle = \frac{1}{2} - \frac{1}{6}\frac{t_xt_y}{t_z}
\end{equation}
However, it is clear from Eq.(19) that $\langle F^1_{optimal}\rangle < \langle F^1_{\max}\rangle$.  Therefore, no general nonunitary trace-preserving reversal channels can do better than unitary ones.

In \cite{Rehacek}, $\chi_{23} = |\Psi^-\rangle_{23}\langle\Psi^-|$ and an imperfect measurement drawn from a one-parametric family of POVMs were considered.  We give here the Kraus operators corresponding to the first element in \cite{Rehacek},
$$
\tilde{\Pi}^{11}_{12} = \frac{\sin\mu}{\sqrt{2}}|11\rangle_{12}\langle 11|
$$
\begin{equation}
\tilde{\Pi}^{12}_{12} = \frac{1}{\sqrt{2}\sqrt{1 + \cos^2\mu}}
(\cos\mu|01\rangle_{12} - |10\rangle_{12})(\cos\mu{_{12}}\langle 01| - {_{12}}\langle 10|)
\end{equation}
In this case, Eq.(6) reduces to
$$
A^{114l} = \sqrt{\frac{1}{p_1}}{_{12}}\langle\tilde{P}_l|
(\tilde{\Pi}^{11}_{12} \otimes I_3)U_{1(2)3}|\tilde{\Psi}^-\rangle_{12},
$$
\begin{equation}
A^{124l} = \sqrt{\frac{1}{p_1}}{_{12}}\langle\tilde{P}_l|
(\tilde{\Pi}^{12}_{12} \otimes I_3)U_{1(2)3}|\tilde{\Psi}^-\rangle_{12},
\end{equation}
where
$$
p_1 = \frac{1}{4}(1 - \sin^2\mu\cos\theta).
$$
Adopting
$$
|\tilde{P}_1\rangle_{12} = |00\rangle_{12},\
|\tilde{P}_2\rangle_{12} = \frac{1}{\sqrt{1 + \cos^2\mu}}
(|01\rangle_{12} + \cos\mu |10\rangle),\
$$
\begin{equation}
|\tilde{P}_3\rangle_{12} = \frac{1}{\sqrt{1 + \cos^2\mu}}
(\cos\mu |01\rangle_{12} - |10\rangle_{12}),\
|\tilde{P}_4\rangle_{12} = |11\rangle_{12},
\end{equation}
we have explicitly
\begin{equation}
A^{1144} = \frac{\sin\mu}{\sqrt{1 - \sin^2\mu\cos\theta}}
\left(\begin{array}{cc} 0 & 1 \\ 0 & 0 \end{array}\right),\
A^{1243} = \frac{1}{\sqrt{1 - \sin^2\mu\cos\theta}}
\left(\begin{array}{cc} \cos\mu & 0 \\ 0 & 1 \end{array}\right),
\end{equation}
a trace-decreasing quantum channel, which has dependence on the input state.  It was shown in \cite{Rehacek} that there are values of $\mu$ where general nonunitary reversal channels are better than unitary ones.

In conclusion, we have demonstrated how to arrive at the generalized depolarizing channel, Eq.(13), by considering a teleportation scheme involving the entangled state, Eq.(8), and joint measurement, Eq.(9).  We then show how the the techniques introduced in Ref. \cite{Rehacek} allow us to determine that unitary transformations optimally reverse the generalized depolarizing channel, Eq.(13).  Same considerations are given to the teleportation scheme in Ref. \cite{Rehacek}, and we arrive at a quantum channel which reveals information about the input state.  In this case, optimal reversal requires more general quantum operations which are not unitary.  In principle, appropriate choices of entangled state $\chi_{23}$ and joint measurement $\tilde{\Pi}^{ij}_{12}$ would yield any desired quantum operation $\cal E$.  The iterative procedure in Ref. \cite{Rehacek} could then be applied to identify and design the desired optimal reversing quantum operation $\cal R$.

The author thanks Yuri Suhov and Andrew Skeen for useful discussions.  This publication is an output from project activity funded by The Cambridge MIT Institute Limited (``CMI'').  CMI is funded in part by the United Kingdom Government.  The activity was carried out for CMI by the University of Cambridge and Massachusetts Institute of Technology.  CMI can accept no responsibility for any information provided or views expressed.

\end{document}